\documentclass{iaus}
\usepackage{graphicx}

\newcommand{\kms}{{\rm km}\,{\rm s}^{-1}}

\newcommand{\msun}{M_\odot}
\newcommand{\mg}{M_{\rm g}}
\newcommand{\ms}{M_*}
\newcommand{\mtot}{M_{\rm tot}}
\newcommand{\mdyn}{M_{\rm dyn}}
\newcommand{\mphot}{M_{\rm phot}}
\newcommand{\msunpc}{M_\odot\,{\rm pc}^{-3}}

\newcommand{\tcr}{t_{\rm cr}}
\newcommand{\trem}{t_{\rm exp}}
\newcommand{\trh}{t_{\rm rh}}
\newcommand{\tdis}{t_{\rm dis}}
\newcommand{\tdisgmc}{t_{\rm dis}^{\rm GMC}}

\newcommand{\rh}{r_{\rm h}}
\newcommand{\rj}{r_{\rm J}}
\newcommand{\reff}{r_{\rm eff}}
\newcommand{\rv}{r_{\rm v}}

\newcommand{\rhoh}{\rho_{\rm h}}
\newcommand{\rhogmc}{\rho_{\rm gas}}
\newcommand{\rhogmcsun}{\rho_{\rm gas}^\odot}
\newcommand{\rhoj}{\rho_{\rm J}}
\newcommand{\sigmaoned}{\sigma_{\rm 1d}}
\newcommand{\sigmagas}{\Sigma_{\rm gas}}
\newcommand{\sigmabin}{\sigma_{\rm bin}}

\newcommand{\vrms}{v_{\rm rms}}
\newcommand{\vesc}{v_{\rm esc}}
\newcommand{\vs}{\vspace{0.1cm}}
\newcommand{\xie}{\xi_{\rm e}}

\title[IAUS 266.~~Star cluster disruption] %% give here short title %%
{Star cluster disruption}

\author[Mark Gieles]   %% give here short author list %%
{Mark Gieles}
\affiliation{European Southern Observatory, Casilla 19001, Santiago, Chile\\
email: {\tt mgieles@eso.org}
}

\pubyear{2009}
\volume{266}  %% insert here IAU Symposium No.
\pagerange{1--12}
\jname{Star Clusters}
\editors{Richard de Grijs \& Jacques R. D. L\'epine, eds.}
\begin{document}

\maketitle
\begin{abstract}
Star clusters are often used as tracers of major star formation events
in external galaxies as they can be studied up to much larger
distances than individual stars.  It is vital to understand their
evolution if they are used to derive, for example, the star formation
history of their host galaxy.  More specifically, we want to know how
cluster lifetimes depend on their environment and structural
properties such as mass and radius. This review presents a theoretical
overview of the early evolution of star clusters and the consequent
long term survival chances. It is suggested that clusters forming with
initial densities of $\gtrsim10^4\,\msunpc$ survive the gas expulsion,
or ``infant mortality", phase. At $\sim10\,$Myr they are bound and
have densities of $\sim10^{3\pm1}\,\msunpc$. After this time they are
stable against expansion by stellar evolution, encounters with giant
molecular clouds and will most likely survive for another Hubble time
if they are in a moderate tidal field. Clusters with lower initial
densities ($\lesssim100\,\msunpc$) will disperse into the field within
a few 10s of Myrs.  Some discussion is provided on how extra galactic
star cluster populations and especially their age distributions can be
used to gain insight in disruption.  \keywords{galaxies: star
  clusters, open clusters and associations: general}
\end{abstract}
\firstsection 

%%%%%%%%%%%
\section{Introduction}
In the Milky Way the majority of star formation occurs in embedded
clusters (\cite[Lada \& Lada 2003]{la03}). At the same time, only a
small fraction of stars in the disc resides in (open) clusters
indicating that most of the clusters/associations are not long lived
and disruption on time-scales of the order of the ages of open
clusters is important.

\begin{figure}[!t]
\begin{center}
\vspace{-0.7cm}
 \includegraphics[width=0.55\textwidth]{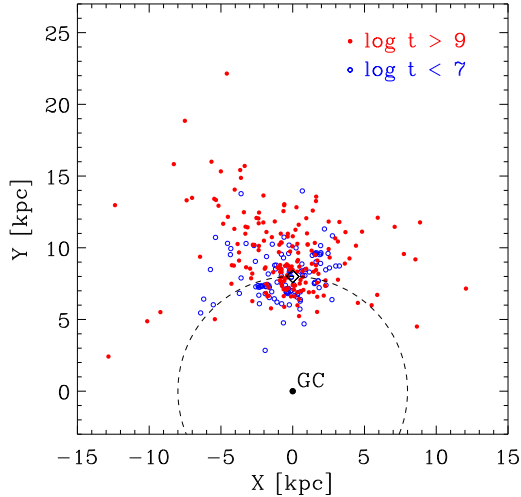} 
 \caption{The distribution of young and old open clusters in the
   Galactic plane, based on the catalogue of \cite[Dias et
     al. (2002)]{di02}. The old open clusters ($\ge1\,$Gyr) are found
   preferentially towards the Galactic anti-centre.}
   \label{fig1}
\end{center}
\end{figure}

Historically, studies on the lifetimes of star clusters focussed on
the open clusters in the Milky Way. In the end of the 50's the
scarcity of old ($\gtrsim\,$Gyr) open clusters was noted nearly
simultaneously in three studies: \cite[van den Bergh (1957)]{vdb57},
\cite[von Hoerner (1958)]{vh58} and \cite[Oort (1958)]{oo58}. It was
already speculated then by these authors that this was due to their
finite lifetimes. The first quantitative explanation for this came
from \cite[Spitzer (1958)]{sp58} who showed that interactions with
Giant Molecular Clouds (GMCs) are a plausible cause for the disruption
of open clusters. In his model a cluster with a density of
$1\,\msunpc$ survives the periodic heating due to passing GMCs for
about 200\,Myr.  \cite[Wielen (1971)]{wi71} constructed an age
distribution of open clusters within 1\,kpc of the sun and derived a
median lifetime of 200\,Myr. Because of this remarkable agreement with
the theoretical predictions by Spitzer, GMC encounters have since then
been held as the dominant disruption mechanism for open
clusters. Another supporting argument for this idea came from the fact
that the old open clusters are strongly concentrated towards the
Galactic anti-centre, where the density of GMCs is low, and their
survival chances high (\cite[van den Bergh \& McClure 1980]{vdb80} and
also Fig.~\ref{fig1}).

Later, \cite[Elson \& Fall (1985)]{ef85} and \cite[Hodge (1987)]{ho87}
constructed age distributions for the LMC and SMC clusters,
respectively. They both compared to Wielen's result and showed that
the age distributions in the Magellanic Clouds extend to older ages,
and consequently concluded that these clusters survive longer than
their Milky Way counterparts.

This review attempts to summarise work done on the disruption of star
clusters since the seminal studies mentioned above.  Three time spans
are considered in which different destruction mechanisms are
important:
\begin{itemize}
\item first few Myr: Expulsion of residual gas from the star formation process
\item first few 100 Myr: Mass loss due to stellar  evolution
\item first Gyr: Interactions with GMCs and 2-body relaxation in a
  tidal field
\end{itemize}
These points will be discussed from a theoretical and observational
viewpoint in \S~\ref{sec:theory} and \S~\ref{sec:obs}, respectively.

%%%%%%%%%%%%%%%%%%
\section{Theoretical considerations}
\label{sec:theory}

%------------------------------------------------------------------------------
\subsection{The first few Myrs: gas expulsion and ``infant mortality'' of clusters}
\label{ssec:im}
A lot of attention has recently gone to the disruption of star
clusters just after their formation due to the removal of residual gas
from the star formation process by the feedback from hot stars. This
was already considered by \cite[Tutukov (1978)]{tu78} and \cite[Lada
  \& Lada (2003)]{ll03} argue it could be the explanation for the
strong drop at a few Myrs in the age distribution of Galactic embedded
and open clusters.  The arguments why this is destructive are as
follows: if the embedded stars are in virial equilibrium with their
surrounding, then the stellar velocities are too high once the gas is
removed. The cluster consequently expands or even complete
dissolves. If we define the total mass of the system as
$\mtot=\mg+\ms$, with $\ms$ the stellar mass and $\mg$ the gas mass,
then virial equilibrium before gas expulsion can be written as
\begin{equation}
\vrms^2=\frac{G\mtot}{2\rv}.
\label{eq:vrms}
\end{equation}
Here $G$ is the gravitational constant, $\vrms$ is the root-mean
square velocity of the stars and $\rv$ is the virial radius of the
(embedded) cluster. The latter relates to the half-mass radius as
$\rv\approx1.25\rh$, slightly dependent on the density profile. Since
$\vrms^2$ is three times the 1-dimensional velocity dispersion
squared, $\sigmaoned^2$, and $\rh$ is $4/3$ times the half-mass radius
in projection, $\reff$, Eq.~\ref{eq:vrms} can also be written in
observable quantities as
\begin{equation}
\sigmaoned^2=\frac{G\mtot}{\eta\reff},
\label{eq:sigmaoned}
\end{equation}
with $\eta\approx10$. In most models it is assumed that the gas
expulsion time-scale, $\trem$, is much shorter than the crossing time
of the stars, $\tcr$. The latter is defined as

\begin{eqnarray}
\tcr&=&2\sqrt{2}\left(\frac{\rv^3}{GM}\right)^{1/2}\\
      &\approx&0.65\,{\rm Myr}\left(\frac{\rhoh}{10^3\,\msunpc}\right)^{-1/2}.
\label{eq:tcr}
\end{eqnarray}
For $\tcr>>\trem$ the positions of the stars are static during the
change of the potential and the cluster's response can be calculated
using the {\it impulsive approximation}.

\begin{figure}[t]
\begin{center}
\vspace{-0.7cm}
 \includegraphics[width=0.55\textwidth]{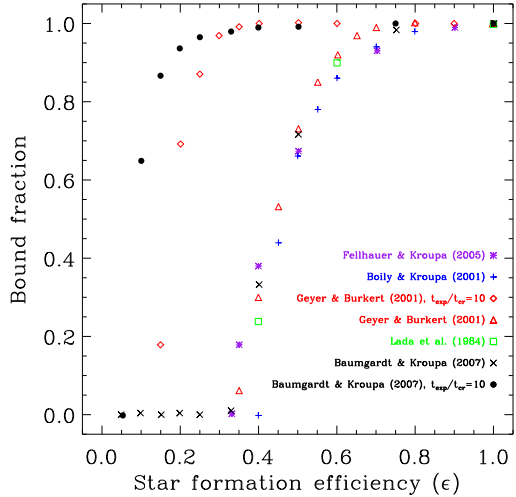} 
 \caption{The fraction of stellar mass that remains bound after gas
   expulsion for different SFEs as found by several authors using
   $N$-body simulations.  Data taken from \cite[Baumgardt \& Kroupa
     (2007)]{bk07}.}
   \label{fig2}
\end{center}
\end{figure}

In here the velocities of the stars right after gas removal are still
defined by the total mass of the embedded cluster (Eq.~\ref{eq:vrms})
and hence the cluster will expand to find a new equilibrium.  Lets
take $\ms=\epsilon\mtot$, such that $\epsilon$ is the star formation
efficiency (SFE).  The final radius can then be expressed in
$\epsilon$ and the initial radius, $\rv(0)$, as (\cite[e.g. Hills
  1980]{hi80})
\begin{equation}
\frac{\rv}{\rv(0)}=\frac{\epsilon}{2\epsilon-1}.
\label{eq:expimp}
\end{equation}
From this it can be seen that for $\epsilon\le0.5$ the cluster
dissolves. Since this value is high for a SFE, gas expulsion is a
plausible cause for the disruption of a large fraction of the embedded
clusters.  A value of $\epsilon=0.5$, however, is for several reasons
an upper limit for complete destruction of the cluster: \vs
\begin{itemize}
\item{\it Escape of unbound stars:} For $\epsilon\le0.5$ the system is
  globally unbound since the total energy (kinetic plus potential) is
  positive. However, most of the positive energy is carried away by
  escaping stars which have velocities much higher than
  $\vrms$. Almost all dynamical models find that for
  $\epsilon\gtrsim0.33$ a fraction of the stars will remain bound (see
  Fig.~\ref{fig2}).  \vs
\item {\it Sub-virial initial conditions:} This was first discussed by
  \cite[Lada et al. (1984)]{la84} who considered a collapsing cloud in
  which the stars are dynamically ``cold" while the gas is removed
  (see also \cite[Proszkow et al. 2009]{pr09}).  If the stellar
  velocities are only a fraction $f$ of the virial velocities then
  Eq.~\ref{eq:expimp} becomes $\rv/\rv(0)=\epsilon/(2\epsilon-f^2)$,
  i.e. the condition for complete disruption becomes $\epsilon=f^2/2$,
  which for $f=0.5$ is $\epsilon=0.125$ (\cite[Goodwin 2008]{go08}),
  much lower than the analytic estimate of $\epsilon=0.5$ for
  initially virialised embedded clusters. The combination of
  $\epsilon$ and the initial virial state, or the {\it effective} SFE
  (\cite[Goodwin \& Bastian 2008]{gb08}), thus determines the survival
  chance.  \vs
\item {\it Clumpy initial conditions:} \cite[Fellhauer et
  al. (2009)]{fe09} showed that if stars form in a clumpy fashion, and
  if these clumps merge  quickly, then the survival chance of the merger
  remnant is enhanced.  \vs
\item {\it Slow gas removal:} If the gas expulsion time-scale is
  longer than $\tcr$, the cluster will expand adiabatically and this
  can result in a bound cluster for $\epsilon$ as low as $0.1$
  (\cite[e.g. Geyer \& Burkert 2001]{gb01}; \cite[Baumgardt \& Kroupa
    2007]{bm07} and Fig.~\ref{fig2}).
\end{itemize}
\vs Therefore, the importance of the gas removal phase in disrupting
star clusters depends on many parameters: the initial density of the
embedded cluster, the gas removal time-scale, the initial virial state
of the embedded cluster, the star formation efficiency and the initial
configuration of the stars (clumpy or smooth density profile). This
has led to several different predictions. For example: most models
assume that gas removal is instantaneous.  If $\epsilon$ is the same
for all clusters then the disruption rate is independent of cluster
mass. \cite[Baumgardt et al. (2008)]{bpk08} advocate, however, that
gas expulsion less efficient in disrupting massive clusters because of
their higher binding energy per unit mass. This makes the general
prediction that the exposed cluster mass function should be shallower
than the embedded cluster mass function (see
\cite[Parmentier~et~al.~2008]{pa08}).
 
Since the embedded phase is short lasting ($\sim1\,$Myr) most of these
parameters are poorly constrained from observations and the
interpretation of the models is therefore sensitive to the assumptions
that are made for the initial conditions. In \S~\ref{sec:obs} possible
observational signatures of this early disruption will be discussed.

%---------------------------------------------------
\subsection{The first few 100 Myrs: mass loss due to stellar evolution }
\label{ssec:sev}
Clusters that survive the gas expulsion phase continuously lose mass
through the evolution of the member stars. For a Kroupa type IMF
between $0.1\,\msun$ and $100\,\msun$ the cluster mass reduces roughly
by 10\%/20\%/30\% in the first 10/100/500\,Myr. Since these stellar
evolution time-scales are much longer than typical values of $\tcr$,
the cluster will in most cases react by an adiabatic expansion. This
expansion is analogous to the increase of the orbital separation of a
binary when one of the members loses mass. It is far less sensitive to
the fractional mass loss than in the impulsive approximation, since
\begin{eqnarray}
\frac{\rv(t)}{\rv(0)}&=&\frac{M(0)}{M(t)}\\
                              &=&\frac{1}{\mu}.
\end{eqnarray}
In the last step $\mu$ is used for the fraction of mass that remains
and this can be compared to $\epsilon$ in the instantaneous mass loss
case (Eq.~\ref{eq:expimp}). For $\mu=0.5$ the cluster will expand by a
factor of two and not lose mass, whereas for impulsive mass loss the
cluster nearly dissolves, as we have seen in \S~\ref{ssec:im}.

The situation is somewhat more complicated than this simple
prediction.  The expansion is more severe and can even result in
complete disruption if the cluster is mass segregated before the bulk
of the stellar evolution takes place (\cite[Vesperini et
  al. 2009]{vmpz09}).  If the cluster is not primordially mass
segregated, it will still segregate while the stars evolve
(\cite[Applegate 1986]{ap86}) and this can eventually lead to an
enhanced expansion at later times. Finally, the fate of the cluster
also depends on the Jacobi (or tidal) radius ($\rj$). The combined
effect of mass loss by stellar evolution and dynamical evolution in a
tidal field was considered by \cite[Fukushige \& Heggie
  (1995)]{fh95}. They show that when clusters expand to a radius of
$\sim0.5\,\rj$ they lose equilibrium and most of their stars
overflow $\rj$ in a few crossing times. As was the case with gas
expulsion, a lot more observational constraints on the initial
conditions of star clusters, such as the degree of primordial mass
segregation and the initial radius with respect to the tidal radius,
are needed before we can make general statements about the effect of
stellar evolution on the evolution of star clusters.

%--------------------------------------------------------------------
\subsection{Interaction with Giant Molecular Clouds}
\label{ssec:gmc}
An external disruptive factor, already considered by
\cite[Spitzer~(1958)]{sp58}, is cluster encounters with Giant
Molecular Clouds (GMCs). Since GMCs are typically more massive than
clusters, the cluster is more affected by an encounter than the cloud
(\cite[Theuns 1991]{th91}). The cluster lifetime due to periodic
heating of passing clouds is inversely proportional to the volume
density of molecular gas, $\rhogmc$, and proportional to the density
of the cluster, a typical result for disruption by external tidal
``shocks" (e.g. \cite[Ostriker et al. 1972]{os72})
\begin{equation}
\tdisgmc\approx1\,{\rm Gyr}
                \left(\frac{\rhogmcsun}{\rhogmc}\right)
                \left(\frac{\rhoh}{10\,\msunpc}\right).
\label{eq:gmc}
\end{equation}
Here $\rhogmcsun\approx0.03\,\msunpc$ is the molecular gas density in
the solar neighbourhood and the constant is taken from \cite[Gieles et
  al. (2006)]{gi06} and equals within a factor of two the original
result by \cite[Spitzer (1958)]{sp58}.  The dependence of $\tdis$ on
$\rhogmc$ means that lifetimes of star clusters also roughly scale
inversely with the observable surface density of molecular gas,
$\sigmagas$ enabling us to make order of magnitude
predictions for the lifetimes of clusters in other galaxies.  In
spiral galaxies disruption by GMC encounters is especially important
in the early stages of evolution since clusters form in the thin
gaseous disc where $\rhogmc$ is high.  Older clusters are typically
more associated with the thick disc where $\rhogmc$ is low and
GMC encounters are less frequent.  Since young ($\lesssim$1\,Gyr)
clusters in spiral galaxies have a near constant radius
(e.g. \cite[Larsen 2004]{la04}), \cite[Gieles et al. (2006)]{gi06}
argue that the dissolution time due to GMC encounters is longer for
more massive clusters, due to their higher density. It is not clear
whether this lack of a mass-radius relation is a universal property
imprinted by formation, or the result of an evolutionary effect. This
point will be addressed in \S~\ref{sec:obs}.

%---------------------------------------------------
\subsection{2-body relaxation in the tidal field of the host galaxy}
On somewhat longer time-scales stars are lost from clusters through
2-body relaxation in which stars get accelerated to the escape
velocity, $\vesc$ (\cite[e.g. Ambartsumian 1938]{am38}; \cite[Spitzer
  1940]{sp40}). The relevant time-scale of escape is, to first order,
the half-mass relaxation time (\cite[Spitzer 1987]{sp87})
\begin{equation}
\trh=0.138\frac{N^{1/2}\rh^{3/2}}{\sqrt{\bar{m}G}\ln\Lambda},
\label{eq:trh}
\end{equation}
where $N$ is the number of stars, $\bar{m}$ is the mean stellar mass
and $\ln\Lambda$ is the Coulomb logarithm: $\ln\Lambda\approx\ln0.11N$
(\cite[Giersz \& Heggie 1994]{gh94}). The relaxation time is
approximately the time stars need to establish a Maxwellian velocity
distribution. A fraction $\xie$ of the stars in the tail of the
distribution have velocities larger than $\vesc$ and consequently
escape. Assuming that this high velocity tail is refilled every $\trh$
then the dissolution time-scale is $\tdis=\trh/\xie$.  For isolated
clusters $\vesc=2\,\vrms$. For a Maxwellian velocity distribution a
fraction $\xie=0.0074$ has $v>2\,\vrms$ and then
$\tdis=137\,\trh$. For tidally limited cluster $\xie$ is higher since
$\vesc$ is lower. For a typical cluster density profile
$\xie\approx0.033$, implying $\tdis\approx30\,\trh$ (\cite[Spitzer
  1987]{sp87}). The escape fraction $\xie$ is often taken constant
(e.g. \cite[Gnedin \& Ostriker 1997]{go97}).  But $\xie$ depends on
$\rh$ (through $\vrms$) and on the strength of the tidal field, or
$\rj$ (through $\vesc$). Effectively, $\xie$ depends on the ratio
$\rh/\rj$ (e.g. \cite[Spitzer \& Chevalier 1973]{sc73}; \cite[Wielen
  1988]{wi88}). \cite[Gieles \& Baumgardt (2008)]{gb08} show that
$\xie\propto(\rh/\rj)^{3/2}$ for $\rh/\rj\gtrsim0.05$ (the tidal
regime). Together with Eq.~\ref{eq:trh} we then find for clusters on
circular orbits in the tidal regime that $\tdis\propto N/\omega$,
apart from the slowly varying Coulomb logarithm. Here $\omega\equiv
V_G/R_G$ is the angular frequency in the galaxy and $R_G$ and $V_G$
are the galactocentric distance and the velocity around the galaxy
centre, respectively.  So for a flat rotation curve we find
$\tdis\propto R_G$ for a cluster of a given mass (e.g. \cite[Chernoff
  \& Weinberg 1990]{cw90}; \cite[Vesperini \& Heggie 1997]{vh97}).
This linear dependence of $\tdis$ on $R_G$ makes it difficult to
explain the universality of the globular cluster mass function from
dynamical evolution of a power-law initial cluster mass function
(e.g. \cite[Vesperini et al. 2003]{ve03}), but this will not be
discussed here. Note that $\omega$ can also be written as
$\omega\propto\rhoj^{1/2}$, where $\rhoj$ is the density within $\rj$,
which is how \cite[Lee \& Ostriker (1987)]{lo87} present it.

\cite[Fukushige \& Heggie (2000)]{fh00} show that the linear scaling
of $\tdis$ with $N$ is slightly affected by the finite time it takes
stars to find the Lagrangian ``exit" points. \cite[Baumgardt
  (2001)]{b01} showed that $\tcr$ is of importance and finds for equal
mass clusters that $\tdis\propto\trh^{3/4}\tcr^{1/4}$. \cite[Baumgardt
  \& Makino (2003)]{bm03} find that this scaling also holds for models
of clusters with a mass spectrum, stellar evolution and for different
types of orbits in a logarithmic potential. Their result for $\tdis$
can be summarised as
\begin{equation}
\tdis\approx2\,{\rm Myr} 
            \left(\frac{N}{\ln\Lambda}\right)^{3/4}\frac{R_G}{{\rm kpc}}
            \left(\frac{V_G}{220\,\kms}\right)^{-1}(1-\varepsilon),
\label{eq:tdis3}
\end{equation}
where $\varepsilon$ is the eccentricity of the orbit. If the
 Coulomb logarithm is taken into account the scaling is
something like $\tdis\propto N^{0.65}$ (\cite[Lamers et
  al. 2005]{la05}).

%%%%%%%%%%%%%%%%%%%%%%
\section{Observational signatures of cluster disruption}
\label{sec:obs}
%----------------------------------------------------------------------------------------
\subsection{The velocity dispersion of young clusters}
\cite[Ho \& Filippenko (1996)]{hf96} were the first to demonstrate
that one can empirically determine the dynamical mass, $\mdyn$, of
star clusters beyond the Local Group using their integrated
light. They used high resolution spectroscopy to measure $\sigmaoned$
from the width of the spectral lines and high angular resolution (HST)
imaging to measure $\reff$. The dynamical mass then follows from the
virial relation (Eq.~\ref{eq:sigmaoned}). They did this for the
$\sim10$\,Myr old super star cluster ``A" in the dwarf starburst
galaxy NGC~1569 and found that its $\mdyn$ is comparable to the masses
of Milky Way globular clusters, concluding that cluster ``A" is a
young globular cluster. Since then this analysis has been applied to
several dozen more clusters with different ages and masses and in
different galaxies (e.g. \cite[Larsen et al. 2004]{la04b};
\cite[Bastian et al. 2006]{ba06}; \cite[McCrady \& Graham 2007]{mc07};
\cite[Mengel et al. 2008]{me08}).

\begin{figure}[!t]
\begin{center}
\includegraphics[width=0.51\textwidth]{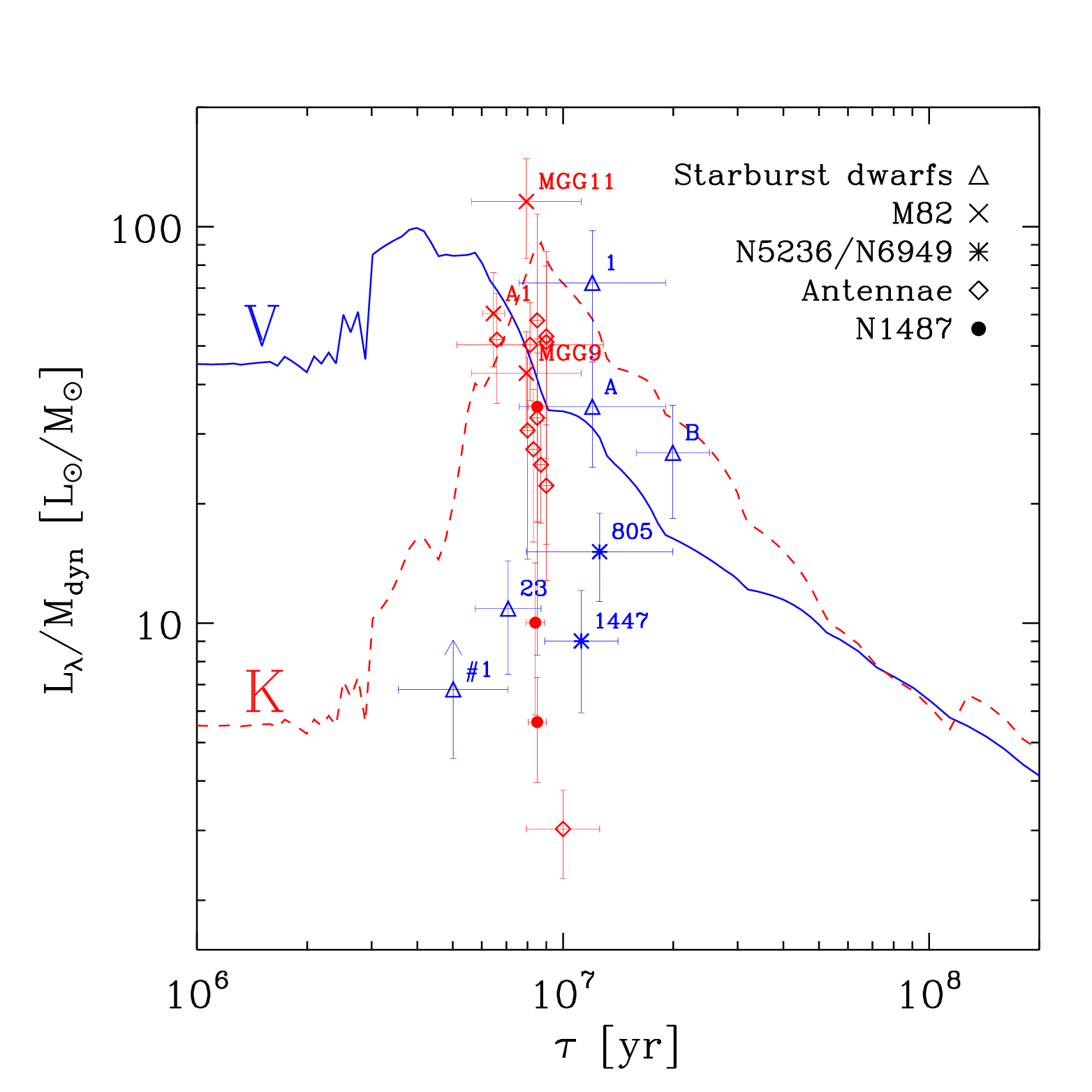} \hspace{-0.65cm}
\includegraphics[width=0.51\textwidth]{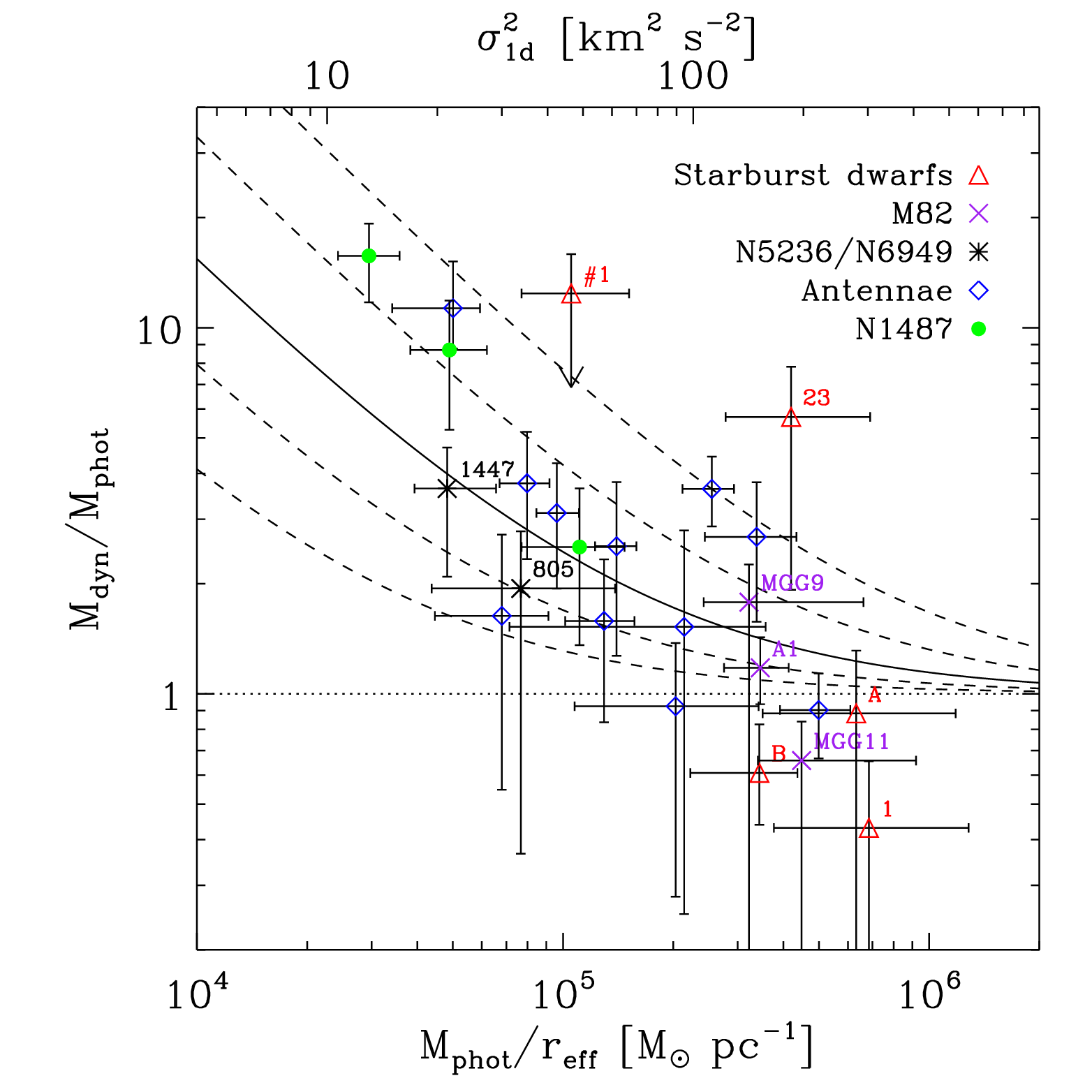}
\end{center}
 \caption{{\it Left:} Light to dynamical mass ratio for 24 clusters
   found in literature. The red and blue points refer to studies done
   in the optical and NIR studies, respectively. Different symbols
   correspond to different galaxy types. The lines show the
   photometric evolution of an SSP model in two filters. {\it Right:}
   Dynamical mass over photometric mass for the same clusters, shown
   as a function of $\mphot/\reff$, which is a proxy of $\sigmaoned$
   (Eq.~\ref{eq:sigmaoned}). The full line is a prediction of the
   effect of binaries on $\mdyn$ (Eq.~\ref{eq:sigbin}), with 1$\sigma$
   and 2$\sigma$ variations due to stochastical fluctuations shown
   as dashed lines (Figure from \cite[Gieles et al. 2009]{gi09b}).
 }
   \label{fig3}
\end{figure}

%----------------------------------------------------------------------------------------
\subsection{Super-virial and dispersing?}
The empirically derived $\mdyn$ relies on the assumption that the
cluster is in virial equilibrium. One way of verifying the validity if
this assumption is by comparing $\mdyn$ to the photometric mass,
$\mphot$. The latter can be derived from the cluster's total
luminosity and age and a comparison to an SSP model. \cite[Bastian et
  al. (2006]{ba06}) provide a compilation of $\mphot$ and $\mdyn$
values of 19 clusters. They find that for the somewhat older clusters
($\gtrsim50-100$\,Myr) there is good agreement between $\mphot$ and
$\mdyn$. For many of the young ($\sim10\,$Myr) star clusters
$\mdyn>>\mphot$.  \cite[Goodwin \& Bastian (2006)]{gb06} suggest that
this is a signature of gas expulsion and the ``infant mortality"
scenario discussed in \S~\ref{ssec:im}, i.e. they conclude that these
clusters are super-virial.

In the left panel of Fig.~\ref{fig3} an updated version of Fig.~5 in
\cite[Bastian et al. (2006)]{ba06} is given. It shows the light to
dynamical mass ratio of 24 young star clusters in different
galaxies. The data are compiled from literature (\cite[McCrady et al.
  2003]{mc03} ; \cite[Bastian et al. 2006]{ba06}; \cite[Smith et
  al. 2006]{sm06}; \cite[McCrady \& Graham 2007]{mc07}; \cite[Moll et
  al. 2007]{mo07}; \cite[{\"O}stlin et al. 2007]{os07}; \cite[Larsen
  et al. 2008]{la08}; \cite[Mengel et al. 2008]{me08}) and will be
discussed in more detail in \cite[Gieles et al. (2009)]{gi09b} and
\cite[Portegies Zwart et al. (2010)]{spz10}. The observations taken in
the visible are shown in blue and the ones taken in the NIR are shown
in red. The photometric evolution in the $V$ and $K$ band from the
\cite[Bruzual \& Charlot (2003)]{bc03} SSP models using a Chabrier IMF
is shown as a full (blue) line and a dashed (red) line,
respectively. All clusters have an age close to $10\,$Myr. This is
because of observational constraints: younger clusters are more
obscured and at 10\,Myr the red supergiants phase sets in making the
cluster brighter (especially in the NIR, see Fig.~\ref{fig3}) and
therefore easier to study.

\begin{figure}[!t]
\begin{center}
\vspace{-0.7cm}
 \includegraphics[width=0.6\textwidth]{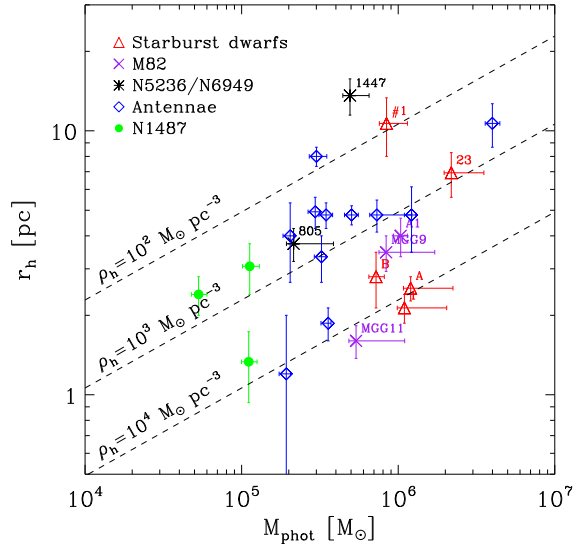} 
\end{center}
 \caption{Mass-radius relation for the same 24 clusters shown in
   Fig.~\ref{fig3}, using the same colour and symbol coding and
   $\rh=(4/3)\reff$. Lines of constant half-mass density,
   $\rhoh\equiv3M/(8\pi\rh^3)$, are overplotted. The cluster densities
   imply $0.2\lesssim\tcr{\rm /Myr}\lesssim2$ (Eq.~\ref{eq:tcr}).}
 \label{fig4}
\end{figure}

It is important to realise what initial conditions are required for
the gas expulsion scenario to work. The models discussed in
\S~\ref{ssec:im} show that after instantaneous gas removal it takes
roughly 10 initial crossing times to either completely disrupt or
expand to a new virial equilibrium. In the models of \cite[Goodwin \&
  Bastian (2006)]{gb06} this corresponds to $\sim25$\,Myr since their
clusters have $\tcr=2.5\,$Myr in the embedded phase. Only with an
initial $\tcr$ of a few Myrs it is possible to ``catch" an unbound
expanding cluster at 10\,Myr. This corresponds to an initial density
(stars and gas) of $\sim100\,\msunpc$ (Eq.~\ref{eq:tcr}). Since the
clusters considered here have densities of $\sim10^3\,\msunpc$ (see
Fig.~\ref{fig4}), gas expulsion can not be responsible for the high
velocity dispersions.  The densities in the embedded phase were at
least a factor $1/\epsilon$ higher because of the gas, but probably
even more since they have also expanded since then. A similar
conclusion can be drawn from the measured velocities, which are of the
order of $20\,\kms$. With such velocities an unbound cluster dissolves
into the field in a few Myrs, incompatible with their radii being only
a few pc. We conclude that the ``infant mortality" phase of dense
clusters only lasts a few Myrs. The clusters considered here have
evolved for at least $10-100\,\tcr$ and are therefore probably bound
objects (an argument also made by McCrady and collaborators).

%----------------------------------------------------------------------------------------
\subsection{Binaries}
The point remains that the measured velocity dispersions are too high
for some of the clusters. It could be that the IMF assumption is
invalid. The virial velocities are higher when the IMF is more bottom
heavy, but the differences between the measured velocities and the
virial velocities are probably too large to be explained by IMF
variations. An alternative is that the velocities are dominated by
orbital motions of binaries in the cluster, not taken into account in
the virial relation (Eq.~\ref{eq:sigmaoned}). \cite[Kouwenhoven \& de
  Grijs (2008)]{kdeg08} considered this and they show that the
$\mdyn/\mphot$ ratio can be elevated by several factors (see also de
Grijs in this contribution). But they concluded that the contribution
of binaries to the velocity dispersion is less then 5\% for clusters
with $\sigmaoned\gtrsim10\,\kms$. So binaries can not be responsible
for the observed velocities according to their model.  But
\cite[Kouwenhoven \& de Grijs (2008)]{kdeg08} did not include the mass
dependent stellar mass-to-light ratio in their models and also did not
take into account stars more massive than $20\,\msun$. So the effect
will be stronger for the clusters considered in Fig.~\ref{fig3}, since
the light is dominated by massive stars ($\sim15\,\msun$) for which
(primordial) multiplicity is high and short periods and mass ratio
close to unity are common.

\cite[Gieles et al. (2009)]{gi09b} derive a simple analytic expression
for the additional velocity dispersion due to orbital motions of stars
in binaries, $\sigmabin$, relative to the dynamical dispersion of a
virialised cluster, $\sigmaoned$

 \begin{equation}
\frac{\sigmabin^2}{\sigmaoned^2} \approx \left(\frac{f}{0.25}\right)
                            \left(\frac{q}{0.6}\right)^{\frac{3}{2}}
                            \left(\frac{m_1}{15\msun}\right)^{\frac{2}{3}}
                            \left(\frac{10^3\,\, {\rm days}}{P}\right)^{\frac{2}{3}}
                            \left(\frac{M/\reff}{10^5\msun\textup{pc}^{-1}}\right)^{-1}.
\label{eq:sigbin}
\end{equation}
Here $f$ is the binary fraction, $q$ is the mass ratio, $m_1$ is the
mass of the primary star (the contribution of secondary star is
ignored) and $P$ is the orbital period. The normalisation constants
are based on properties of massive binaries in the Milky Way and
appropriate for an age of $10\,$Myr. Since the cluster light is
dominated by the most massive stars the measured velocity dispersion
is affected by binaries according to this relation.  The prediction
for $\mdyn$ including the velocities of binaries
($\mdyn\propto\sigmaoned^2+\sigmabin^2$) is shown in the right panel
of Fig.~\ref{fig3} (full line). The dashed lines show the 1$\sigma$
and 2$\sigma$ spread in the model due to stochastics. The discrepancy
between $\mdyn$ and $\mphot$ can largely be explained by virialised
clusters with a modest binary population. Assuming that all clusters
host a similar binary population, the clusters with a low dynamical
dispersion, for which we use $\mphot/\reff$ as a proxy
(Eq.~\ref{eq:sigmaoned}), are affected most. This could be the reason
why the data show a decreasing $\mdyn/\mphot$ ratio for an increasing
$\mphot/\reff$ ratio.

This suggest that these clusters are survivors of the gas expulsion
phase and are at the beginning of the stellar evolution phase.  In the
next Hubble time these clusters will adiabatically expand by a factor
of $\sim2$ due to mass loss by stellar evolution if they are not mass
segregated (\S~\ref{ssec:sev}) and it remains to conclude that these
clusters are good young globular cluster candidates if the tidal field
they are in is not too strong.

%------------------------------------------------------------------------------
\subsection{The mass radius relation of star clusters}
\label{sec:mr}
Several studies have pointed out that there is a very weak dependence
of cluster radius on mass/luminosity (e.g. \cite[Zepf et
  al. 1999]{ze99}; \cite[Larsen 2004]{04}; \cite[Bastian et
  al. 2005]{ba05}). For the clusters in Fig.~\ref{fig4} this seems not
to be the case. Although these are only 24 clusters, and it is far
from a tight relation, these objects seem more consistent with a
constant density of $\rhoh\approx10^{3\pm1}\,\msun$, or $\rh\approx
(M/10^4\,\msun)^{1/3}$ within a factor of two. The difference between
this figure and the studies mentioned above is that the clusters
considered here are confined to a narrow (young) age range. The
mass-radius relation is of importance for disruption since
$\tdisgmc\propto\rhoh$ (\S~\ref{ssec:gmc}). So for a constant cluster
density GMCs will destroy all clusters equally fast.  If the
mass-radius relation evolves in time, for example to a near constant
radius as found by several authors, then $\tdisgmc$ will become mass
dependent. In any case, the densities of these clusters imply that
$\tdisgmc$ is much longer than a Hubble time, even when $\rhogmc$ is
10 times higher than in the solar neighbourhood (Eq.~\ref{eq:gmc}), so
GMC encounters will not play an important role in the disruption of
these clusters at this stage.

In a solar neighbourhood type tidal field the clusters in
Fig.~\ref{fig4} have $\rh/\rj\approx0.03$, within a factor of two true
for all of them. This implies that these clusters are probably also
very stable against the expansion due to mass loss by stellar
evolution, since in \S~\ref{ssec:sev} it was argued that rapid
dissolution due to mass loss by stellar evolution only occurs when
$\rh/\rj$ increases to $\sim0.5$. A similar idea was recently put
forward by \cite[Pfalzner (2009)]{pf09} who discovered two
evolutionary sequences in young Galactic star clusters. A dense group
that starts with a density of $\sim10^5\,\msunpc$ at a few Myr
containing the Arches cluster, NGC~3603 and Trumpler 14, to which they
refer as ``starburst clusters".  They seem to expand along a sequence
of constant mass, where at $10-20\,$Myr there are the red supergiant
clusters (RSGC01 and RSGC02) with a density of
$\sim10^3\,\msunpc$. The second sequence also starts at a few Myr, but
with much lower densities ($\sim10\,\msunpc$) and expands along a
$M\propto1/\rh$ sequence to densities comparable to the field star
density. These are referred to as ``leaky clusters". The clusters in
Fig.~\ref{fig4} could be compared to the red supergiant clusters in
the Milky Way, i.e. the end stage of the starburst sequence.

%%%%%%%%%%%%%%%%%%%%
\section{Discussion}

\begin{figure}[!t]
\begin{center}
\vspace{-0.25cm}
 \includegraphics[width=0.55\textwidth]{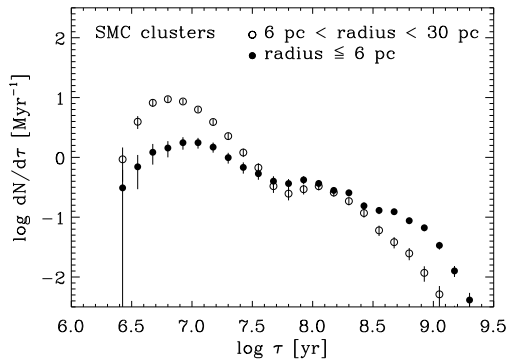} 
\end{center}
 \caption{Age distribution of SMC clusters based on the catalogue of
   \cite[Chiosi et al. (2006)]{chi06}. The sample is split in small
   and the large clusters/associations, with the boundary at 6 pc. The
   histograms are made using a 0.5 dex bin width with different
   starting values (boxcar averaging).}
   \label{fig5}
\end{figure}
\vspace{-0.1cm} Up to the 90s the open clusters in the Milky Way and
the populous clusters in the Magellanic Clouds were the targets of
studies on disruption. HST has enabled us to establish the properties
of larger populations of clusters containing more massive clusters in
quiescent spiral galaxies (e.g. \cite[Larsen \& Richtler 2000]{lr00}),
interacting galaxies (e.g.~\cite[Bastian et al. 2005]{ba05};
\cite[Whitmore et al. 1999]{wh99}) and merger remnants
(e.g. \cite[Miller et al. 1997]{mi97}).

The primary tool used for studies on the disruption of clusters is the
age distribution. Different groups give varying weights to the various
aspects described in \S~\ref{sec:theory} in the interpretation of the
results. There are two flavours of empirically based cluster
disruption models: one is based on {\it externally driven cluster
  disruption}. In this model the dissolution time depends on the mass
of the cluster and its environment. It was based on the age and mass
distributions of luminosity limited cluster samples in different
galaxies (\cite[Boutloukos \& Lamers 2003]{bl03}). The variations in
dissolution time-scales are explained by differences in the tidal
field strength (\cite[Lamers et al. 2005]{la05}) and the GMC density
(\cite[Gieles et al. 2006]{gi06}). The second model assumes that {\it
  internally driven cluster disruption} is most important.  In this
model roughly 80-90\% of the number of clusters is destroyed each age
dex resulting in a (mass limited) age distribution that declines
roughly as $\tau^{-1}$. In this model it is assumed that the infant
mortality of clusters proceeds for a few 100\,Myr and that the
disruption rate is equally fast for all masses. This model is based on
the cluster population of the Antennae galaxies (\cite[Fall et
  al. 2005]{fa05}; \cite[Whitmore et al. 2007]{wh07}) and it was
argued by these authors that their model is universal and should be
able to describe the age distributions of cluster populations in other
galaxies as well.

The universal cluster dissolution scenario has led to some
controversy. \cite[Chandar et al. (2006)]{cha06} show that the age
distribution of SMC clusters declines as $\tau^{-0.85}$, almost the
same as in the Antennae. \cite[Gieles et al. (2007)]{gi07} used the
same data set and show that the age distribution of massive clusters
($\gtrsim10^{3.5}\,\msun$) is flat in the first few 100\,Myr.  They
reproduce the $\tau^{-0.85}$ result for the full sample and conclude
that the decline is caused by detection incompleteness, confirmed by
\cite[de Grijs \& Goodwin (2008)]{gg08} using similar arguments.
\cite[Boutloukos \& Lamers (2003)]{bl03} show that for a constant
formation history and no disruption, the age distribution of a
luminosity limited sample declines as $\tau^{-\zeta(1-\alpha)}$, where
$-\alpha$ is the index of the cluster initial mass function.  The
index $\zeta$ describes the cluster fading with age:
$F_\lambda(\tau)\propto\tau^{-\zeta}$, with $F_\lambda$ being the flux
at wavelength $\lambda$ of a cluster with constant mass. For the
$U(V)$-band $\zeta\approx1.0(0.7)$, resulting in $\tau^{-\eta}$, with
$0.7\lesssim\eta\lesssim1.0$ due to fading alone when $\alpha=2$. If
the cluster mass function is a Schechter function (\cite[Gieles
  2009]{gi09a}; \cite[Larsen 2009]{la09}), the age distribution of a
luminosity limited sample not affected by disruption is as steep as
$0.9\lesssim\eta\lesssim1.4$ (see also Konstantopoulos in this
contribution). Both models rely on the assumption that the cluster
formation history has been constant in the age range
considered. \cite[Bastian et al. (2009)]{ba09} recently pointed out
that this is invalid for interacting galaxies like the Antennae
galaxies and that a recent burst can be confused by mass independent
cluster disruption.

Another issue, especially important for distant cluster populations,
is what we actually define as a star cluster. Fig.~\ref{fig5} shows
the age distribution of clusters and associations in the central
region of the SMC (\cite[Chiosi et al.  2006]{chi06}). The sample is
split in two sub-samples of $\sim200$ clusters, based on their size.
The age distribution of large clusters falls off much quicker than the
one of the compact group. The median radius of both samples are 4.5 pc
and 9 pc, respectively. At a distance of 20\,Mpc it will be very hard
to tell these two groups apart. A speculative suggestion is that more
of the short-lived ``associations" or ``leaky clusters" are included
in cluster samples in distant galaxies, that would not be considered
as genuine star clusters if they could be resolved.

\begin{acknowledgement}
I thank Brad Whitmore and Iraklis Konstantopoulos for an interesting
discussion over lunch at Copacabana beach during this conference and
the organisers of IAUS~266 for an interesting meeting!
\end{acknowledgement}

%\begin{discussion}
%\end{discussion}

\end{document}